\documentclass[epsf]{aa}
\usepackage{natbib}
\usepackage{graphicx}

\def\NewAR{New Astron. Rev.}
\def\ASPConf#1#2{ASP Conf. Ser. #1, #2}
\def\PublisherASP{San Francisco: ASP}
\def\inpress{in press}
\def\astroph#1{ (astro-ph/#1)}

\begin{document}

\title{Deep Transient Optical Fading in the WC9 Star WR 106}
\subtitle{}
\authorrunning{T. Kato et al.}
\titlerunning{Transient Optical Fading in WR 106}

\author{Taichi Kato\inst{1}
        \and Katsumi Haseda\inst{2}
        \and Kesao Takamizawa\inst{3}
        \and Hitoshi Yamaoka\inst{4}
}

\institute{
  Department of Astronomy, Kyoto University, Kyoto 606-8502, Japan
  \and Variable Star Observers League in Japan (VSOLJ), 2-7-10 Fujimidai,
       Toyohashi City, Aichi 441-8135, Japan
  \and Variable Star Observers League in Japan (VSOLJ), 65-1 Oohinata,
       Saku-machi, Nagano 384-0502, Japan
  \and Faculty of Science, Kyushu University, Fukuoka 810-8560, Japan
}

\offprints{Taichi Kato, \\ e-mail: tkato@kusastro.kyoto-u.ac.jp}

\date{Received / accepted }

\abstract{
   We discovered that the WR9-type star WR 106 (HDE 313643) underwent
a deep episodic fading in 2000.  The depth of the fading
($\Delta V \sim$ 2.9 mag)
surpassed those of all known similar ``eclipse-like" fadings in WR stars.
This fading episode was likely to be produced by a line-of-sight
episodic dust formation rather than a periodic enhancement of dust
production in the WR-star wind during the passage of the companion star
though an elliptical orbit.  The overall 2000 episode was composed of
at least two distinct fadings.  These individual fadings seem to more
support that the initial dust formation triggered a second dust formation,
or that the two independent dust formations occurred by the same
triggering mechanism rather than a stepwise dust formation.
We also discuss on phenomenological similarity of the present fading with
the double fading of R CrB observed in 1999--2000.
}

\maketitle

\section{Introduction}

   Wolf--Rayet (WR) stars are massive, luminous stars which have blown
away the hydrogen envelope, and are considered to be immediate precursors
of some kinds of supernovae.  WCL-type stars are a carbon-rich, late-type
subclass of WR stars [for the definition of the subclasses of WC-type stars,
see e.g. \citet{smi90}; more comprehensive information of WR stars can
be found in the catalogue by \citet{vdh01WRcatalog}.
WC9-type stars are the coolest WCL-type stars
which are characterized by strong C\textsc{III} and C\textsc{II} lines,
and the weak or absent O\textsc{V} feature \citep{tor84WC9}.
WC9-type stars have been receiving much astrophysical attention in that
they are one of the most effectively dust-producing environments in
stellar systems (for recent reviews, see \cite{wil95,wil97}).

   The dust-forming process in WCL-type (especially in WC9-type) stars
is known to be either continuous or episodic.  The best-known continuous
dust producer is a binary WR 104 (WC9+B0.5V), renowned for its ``dusty
pinwheel nebula" \citep{tut99,tut02}.  Recently discovered large-amplitude
optical variability even suggests the presence of a continuous ``dust jet"
in the direction of the rotation axis \citep{kat02wr104}.
WR 112 (WC9+OB?) has been also suspected to have a similar dusty pinwheel
\citep{mar02}.

\begin{figure}
  \begin{center}
  \includegraphics[angle=0,width=4.1cm]{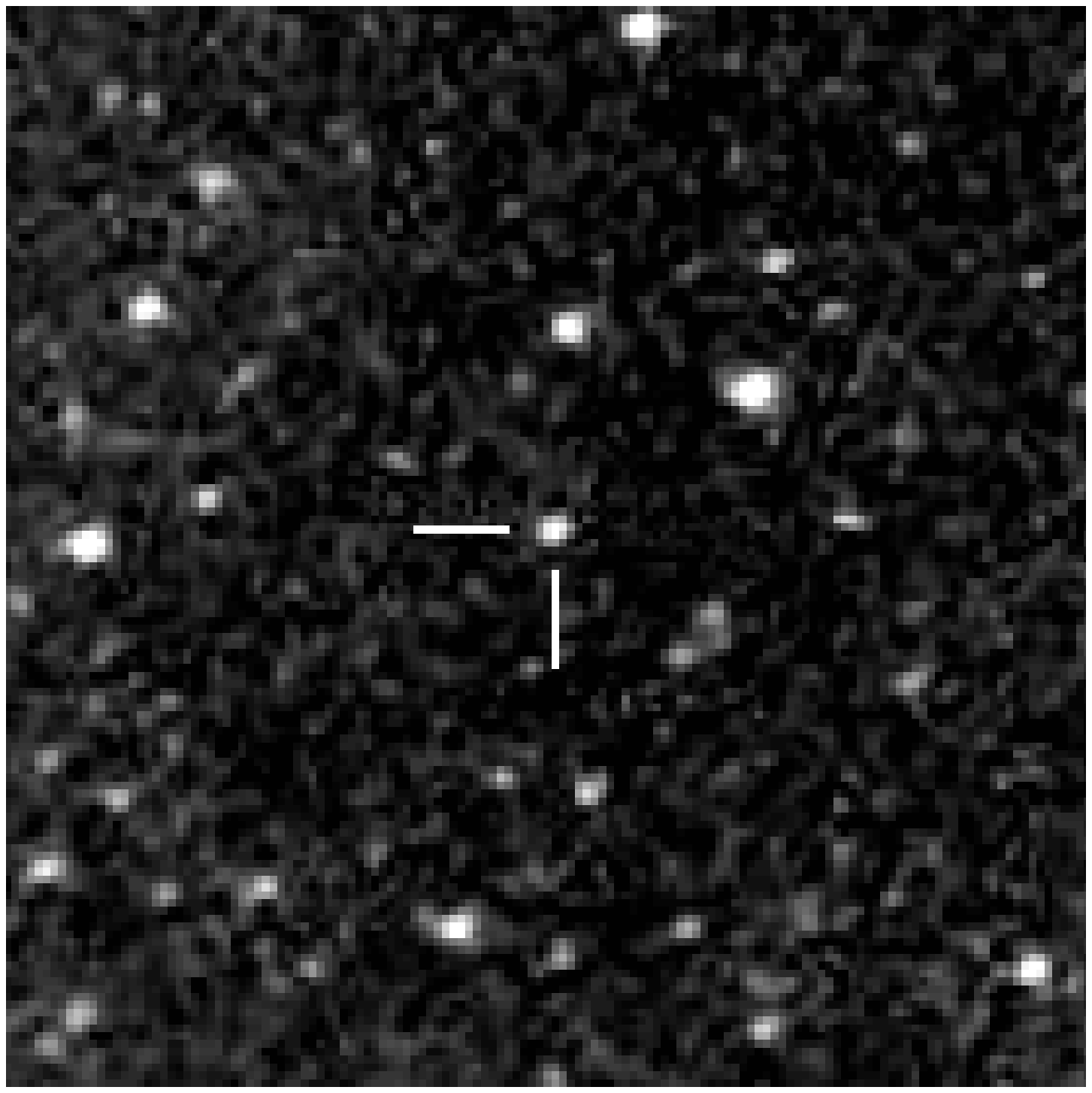}
  \includegraphics[angle=0,width=4.1cm]{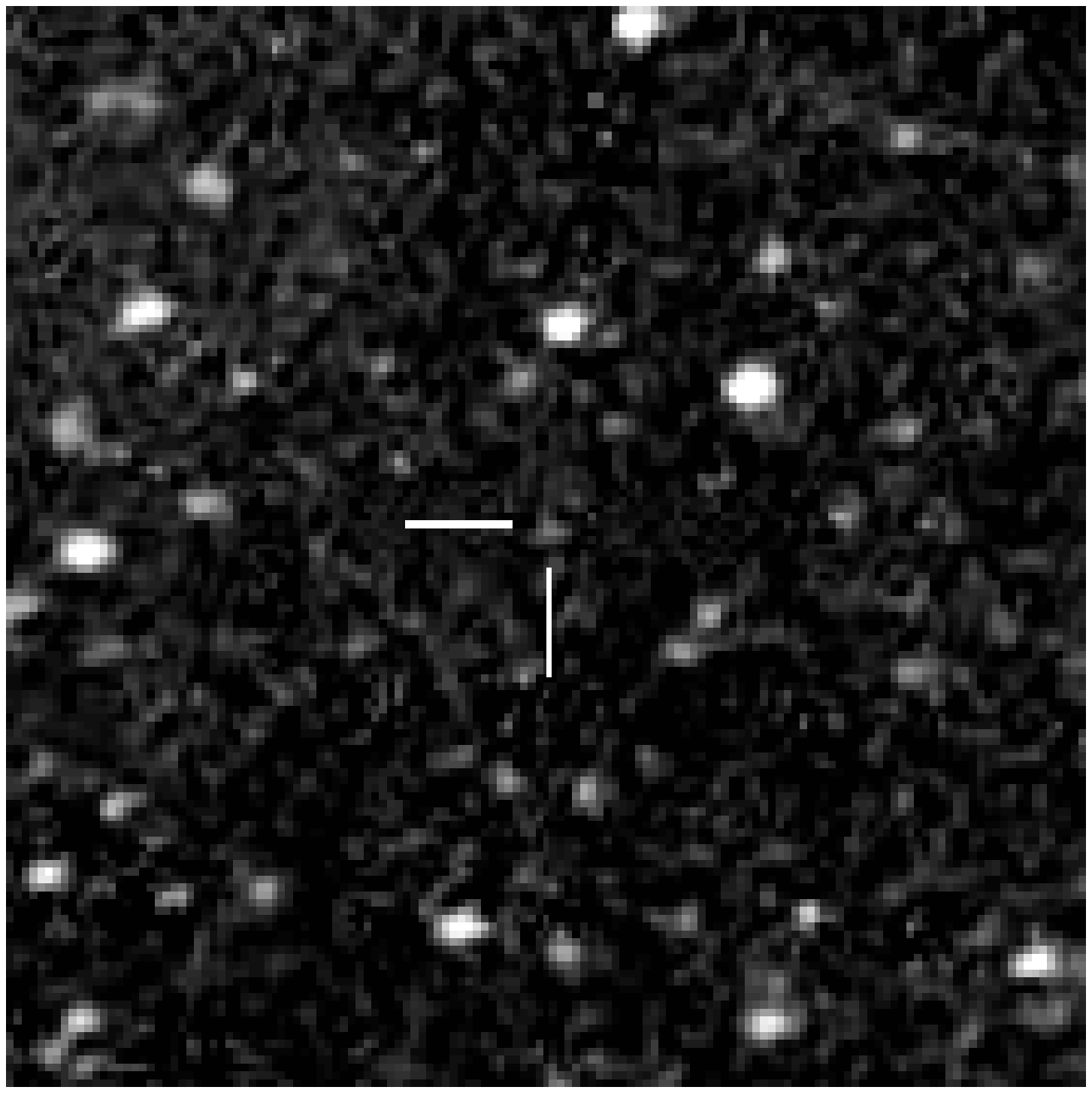}
  \end{center}
  \caption{Variation of WR 106 = Had~V84, recorded with photographs
  taken by one of the authors (KH).
  Each panel shows 10 arcmin square, north is up and east is left.
  The left and right panels were taken on 2000 Aug. 22 and 2000 Apr. 28,
  when the object was at 12.0 mag and 14.1 mag, respectively.
  Such dramatic variability of a Wolf--Rayet star is quite exceptional.}
  \label{fig:image}
\end{figure}

\begin{figure*}
  \begin{center}
  \includegraphics[angle=0,height=5.2cm]{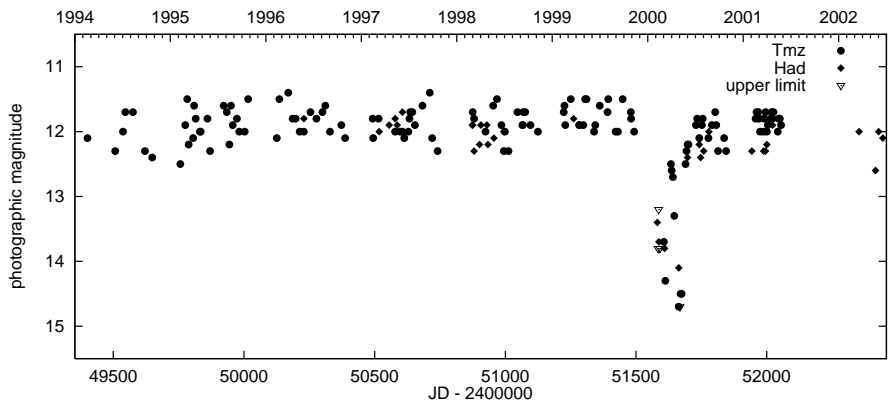}
  \end{center}
  \caption{Light curve of WR 106.  The filled circles and squares represent
  observations by Takamizawa (Tmz) and Haseda (Had), respectively.
  The open triangles represent the upper limits.  The most prominent
  fading was recorded in early 2000.}
  \label{fig:lc}
\end{figure*}

   Another class of manifestation of dust production in WCL stars is
episodic optical fading \citep{veen97} or episodic infrared brightening
\citep{wil90}, which are considered to arise from temporary condensations
of dust clouds.

   In 2001 April, one of the authors (KH) serendipitously discovered a new
variable star named Had~V84 (vsnet-alert 5856),\footnote{
  http://www.kusastro.kyoto-u.ac.jp/vsnet/alert5000/\\msg00856.html.
} which was subsequently identified with WR 106 = HDE 313643
(Fig. \ref{fig:image}).
WR 106 is known to show a strong infrared excess
\citep{coh78IRspecphot,coh95IRASWC,kwo97IRASLRS,pit83IRphot},
which indicates substantial dust formation.
We also noticed that the object was listed as No. 15357 in \citet{fit73},
who suspected 0.13 mag $V$-band variability based an analysis of past
photoelectric archival data.  The object was given a name for suspected
variable star (NSV 10152), but the variability was not confirmed
at that time.

\section{Observation and Results}

   A total of 177 observations were made between 1994 February 17 and
2002 June 18, with twin patrol cameras equipped with a $D$ = 10 cm f/4.0
telephoto lens and unfiltered T-Max 400 emulsions, located at two sites in
Toyohashi, Aichi (KH) and Saku, Nagano (KT).  The passband of observations
covers the range of 400--650 nm.  Photographic photometry was performed
using neighboring comparison stars, whose $V$-magnitudes were calibrated
by T. Watanabe.  The magnitudes were derived by a combination of image
size and density.
The overall uncertainty of the calibration and individual
photometric estimates is 0.2--0.3 mag, which will not affect the following
analysis.  A scatter around the maximum light likely comes from statistical
distribution of errors, although superposed intrinsic variations cannot be
ruled out.

   The resultant light curve is presented in Fig. \ref{fig:lc}.  The star
showed an overall range of variability of between 11.4 and fainter than 14.7
mag.  Taking into measurement errors into consideration, the minimum full
amplitude of the variation is 2.9 mag.

\begin{figure}
  \begin{center}
  \includegraphics[angle=0,height=4.8cm]{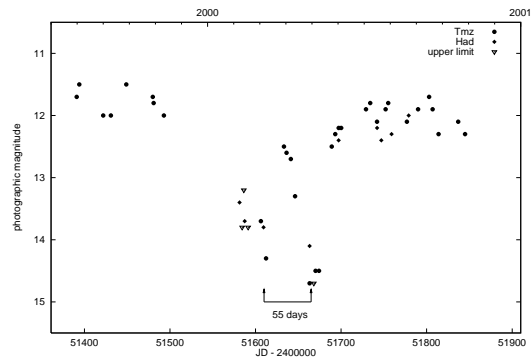}
  \end{center}
  \caption{Enlarged light curve of the 2000 fading episode.  The episode
  is composed of at least two fadings.  The two most prominent fadings
  was separated by $\sim$55 d.}
  \label{fig:fading}
\end{figure}

   Fig. \ref{fig:fading} shows an enlarged light curve of the 2000
fading episode.  This figure clearly demonstrates that the overall
fading episode was composed of at least two distinct fadings.
The two most prominent fadings were separated by $\sim$55 d.
In 2002 May, KH reported a marginal detection of another shallower
($\Delta V \sim$ 0.6 mag).

   An inspection of the available archived images at the USNOFS pixel
server, 9 epochs during 1950-1996, has revealed no distinct fading of WR 106,
suggesting that fadings are rather rare.

\section{Discussion}

   WR 106 was studied for binarity by \citet{wil00companion}.  The lack of
evidence for a companion and the apparent lack of photometric periodicity
(Fig. \ref{fig:lc}) less favor the interpretation of a periodic enhancement
of dust production in the WR-star wind during the passage of the companion
star though an elliptical orbit, as has been proposed in WR 140 \citep{wil90}
and presumably WR 137 \citep{wil01}.

   The present phenomenon seems to be better understood as
an ``eclipse-like", line-of-sight dust formation as proposed
by \citet{veen97}.
The depth of the present phenomenon, however, far surpasses those (up to
1.2 mag in visual wavelengths) of the previously known similar phenomena
in other stars.
Following the interpretation by \citet{veen97}, the production rate of
the optical depth or the dust production rate in the present episode
should be at least a few times larger than in the previously recorded
phenomena.  Furthermore, the observed depth severely constrains the
amount of the unobscured scattered light to be less than 7 \%.

   The present phenomenon is composed of at least two distinct fadings
(Fig. \ref{fig:fading}).  \citet{veen97} reported the presence of
two-step fadings in some fadings.  \citet{veen97} suggested several
possibilities to explain the two-step fadings: (1) sudden enhancement
of the dust production in response to an inflow of additional matter
to the dust production area, (2) non-radial expansion of a neighboring
cloud, or (3) formation of the second cloud in the shade of the first
cloud.  In the present case, the close occurrence of two rare fadings
suggests that they are not a chance superposition of two independent
phenomena, but are more physically related.
The similar observed depths and durations of the two fadings do not
seem to support a stepwise formation of the dust cloud, as represented
by the possibilities (1) and (3).  The present observation seems to more
support that the initial dust formation somehow triggered a second dust
formation in the proximity, or that the two independent dust formations
occurred by the same triggering mechanism.

\begin{figure}
  \begin{center}
  \includegraphics[angle=0,height=4.8cm]{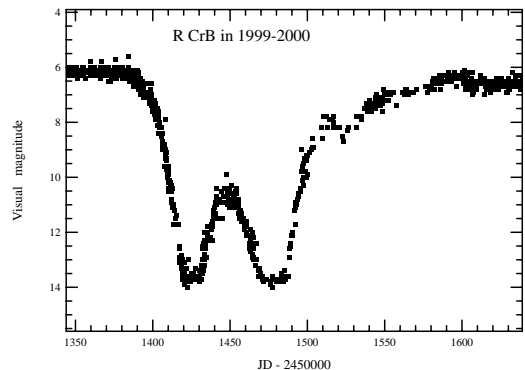}
  \end{center}
  \caption{Light curve of the ``double fading" of R CrB in 1999--2000.
  The data are from VSNET.}
  \label{fig:rcrb}
\end{figure}

   We also note phenomenological similarity of the present fading with
the ``double fading" of R CrB observed in 1999--2000 (Fig. \ref{fig:rcrb},
the data are from VSNET.\footnote{http://www.kusastro.kyoto-u.ac.jp/vsnet/.}).
The fading mechanism proposed by \citet{veen97} being analogous to
the fading mechanism of R CrB stars (for a review, see \cite{cla96rcrb}),
the analogy may suggest a common underlying dust production mechanism
between R CrB stars and WR 106.  Similar double fadings are also known
in some [WC] stars (CPD$-$56$^{\circ}$8032 = He3$-$1333 =
V837 Ara: \cite{pol92WC11phot}; V348 Sgr: \cite{hec85v348sgrphot}), which
are sometimes considered to be related to R CrB-type stars.
It is widely believed that the dust
formation in R CrB stars are associated with pulsation \citep{cla96rcrb}.
Although the large difference of the gravity and temperature between
WR stars and R CrB stars may make it difficult to directly apply the
R CrB-type dust formation to a WR star, a pulsation-type instability
similar to that of R CrB stars in the outer WR wind may have caused
a similar sequence of fadings in a WR star.

\vskip 3mm

The authors are grateful to the observers who reported visual observations
of R CrB to VSNET.
This work is partly supported by a grant-in aid [13640239 (TK),
14740131 (HY)] from the Japanese Ministry of Education, Culture, Sports,
Science and Technology.
This research has made use of the Digitized Sky Survey producted by STScI, 
the ESO Skycat tool, and the VizieR catalogue access tool.
This research has made use of the USNOFS Image and Catalogue Archive
operated by the United States Naval Observatory, Flagstaff Station
(http://www.nofs.navy.mil/data/fchpix/).

\end{document}